\documentstyle[11pt,IAU207_pasp,twoside,psfig]{article}
\def\edcomment#1{\iffalse\marginpar{\raggedright\sl#1\/}\else\relax\fi}
\reversemarginpar

\begin{document}
\title{Old and new tools for understanding the evolution of stars in
clusters }
 \author{Francesca D'Antona}
\affil{Osservatorio Astronomico di Roma, I-00040 Monteporzio (Italy)}

\begin{abstract}
The uncertainties in the modelling of some important inputs of stellar
evolution must be taken into account for a correct interpretation, both of
the HR diagrams of individual stars, and of the integrated colors, of stellar
clusters. After a short discussion of Globular Cluster ages, we focus on the
problem of convection and discuss the parametrization of convection
efficiency and overshooting.

Convective efficiency affects the reliability of the giants colors,
especially for high metallicity: age -- metallicity relations found from
integrated colors of clusters must be regarded with caution.

Non--instantaneous mixing, both in the formal convective region and in the
overshooting region must be taken into account: this is necessary to compute
nucleosynthesis in `hot bottom burning' envelopes, but it may also affect
the color distribution in clump (core helium burning) stars, as shown here
for the case of the LMC cluster NGC 1866.
\end{abstract}
\section{Introduction}
In a Symposium dedicated mostly to the observations of Extragalactic Star
Clusters it is highly necessary to go back to the basics of our knowledge of
stellar evolution, to assess fairly the uncertainties in the derivation of
the physical most interesting parameters of clusters, first from the stars
distribution in the HR diagram, and then from the few integrated quantities
we have available for the farther objects.

The study of stellar evolution is certainly one of the most advanced
branches of astrophysics, and a great progress has been achieved in the
latest years on all the problems of ``microphysics", that is the opacities,
equation of state (EOS), nuclear reaction rates and so forth. Much of this
progress has been forced by the necessity of fitting properly the very
detailed properties of the seismic Sun. For this purpose, e.g., the Livermore
opacity project (Rogers and Iglesias 1992) was developped. Not many have
noticed that the EOS built up for the computation of the Livermore opacities
(Rogers, Swenson \& Iglesias 1996) has resulted to be very important for
modelling both turnoff stars (Chaboyer and Kim 1995) and Horizontal Branch
(HB) stars (Mazzitelli, D'Antona \& Caloi 1995, Caloi, D'Antona \&
Mazzitelli) in Globular Clusters (GCs), contributing to decrease the age of
the most metal poor GCs, the `stellar' age of the Universe. Nevertheless,
uncertainty on this result is still quite large (section 2), as a very high
degree of precision in the models is required to reduce it.

On the other hand, there are problems in the ``macrophysics" of stellar
structure, whose description does not yet come from first principles, but
from parametrizations based on the comparison with observations: the most
important are mass loss and turbulent convection. I will discuss here
this latter problem, in its different aspects (Sect. 3 and 4) .
\section{The age of Galactic Globular Clusters}
\label{gc}
This is an issue in which microphysics plays a complex role.
Good reviews are available (Vandenberg, Stetson and Bolte
1996, Chaboyer 2000, D'Antona 2000a), together with up to date discussions
of the fits with modern stellar models (e.g. Vandenberg 2000).
I wish to remember here that the {\it details} of the
theory determine the precise absolute ages of the galactic Globular Cluster
(GC) system. The age is given by the absolute luminosity of the turnoff of
the GC, so we must very precisely know the distance scale. All the methods
which can be employed to derive it (main sequence (MS) fit to the subdwarfs,
MS, HB and/or RR Lyrae fit to theoretical models,
fit to the White Dwarf sequence or to the lower MS) end up with uncertainties
$\sim 0.25$mag in the turnoff location. As the GC age scales by 1Gyr
for a variation of $\sim 0.07$mag in the turnoff magnitude, the global
uncertainty due to the distance scale is still at least 3Gyr.
Other uncertainties in the microphysics (e.g. the role of gravitational
settling) add another $\pm 1.5$Gyr, so that the absolute
age must still be considered to be $\simeq 13 \pm 3$Gyr. I add that
the {\sl most recent} age determinations all seem to converge towards the
younger ages, and my own favourite bet is 12--14Gyr for the oldest GCs, but
this is certainly not the final.
\section{Convection model: its efficiency and the colors of giants}
\label{conv}
Turbulent convection has two main aspects: the computation of the
{\sl convective fluxes} and the problem of {\sl chemical mixing} into stars.
We need a model to compute the convective fluxes. There are presently three
widely adopted ways of computing superadiabatic
convection in general purpose stellar evolution codes:
1) the Mixing Length Theory (MLT) by B\"ohm Vitense (1958);
2) the Full Spectrum of Turbulence (FST) model by Canuto \& Mazzitelli 1991;
3) the Large Eddy Simulations (LES).
Limits and successes of these formulations are discussed, e.g., by
Canuto \& Christensen Dalsgaard (1998), Mazzitelli (1999), D'Antona (2000b).
Remember that the LES are in principle able to provide a {\it non local}
description of turbulence, but the reliability of their predictions is still
limited by the present computer power. The MLT and FST description both
require {\it parametric} formulations for the extra-mixing or overshooting
problem.
It is well known, but not generally appreciated enough, that the efficiency
of convection determines the stellar $T_{\rm eff}$\ for stars having
important external convective layers. In the Red Giant Branch (RGB) in
particular, the low densities in the envelope imply that convection is
largely overadiabatic. The lower is convection efficiency, the larger is the
actual temperature gradient in the overadiabatic layers, and the smaller is
the resulting stellar $T_{\rm eff}$ (or the redder are the colors). As
example, the MLT efficiency is larger the larger is the parameter $\alpha$,
defined as the ratio between the mixing length and the pressure scale heigth.
A precise value of this parameter can be derived by imposing the fit of the
solar radius at the solar age, but there is no physical reason why convection
should have the same global efficiency in stars different from the present
Sun. So, the smaller is $\alpha$, the lower is the RGB $T_{\rm eff}$.
By itself, the parametrization of convection provides an uncertainty by
several hundred degrees in the location of the RGB, for a given evolving mass
and metallicity (Z).
As we know, the RGB shifts to cooler T$_{\rm eff}$'s by increasing Z (as the
opacity increases), and to hotter T$_{\rm eff}$'s by decreasing the age (as
the evolving mass increases). Uncertainty in the calibration color versus Z
may lead to uncontrolled feedbacks on our interpretation of observations. For
example, we may be led to attribute a younger age to a metal rich cluster,
simply as an artifact of its integrated color being bluer than the color we
expect on the basis of our adopted models. For understanding the HR diagram
location of single stars, and to derive information from integrated colors,
we must know exceedingly well the functional dependence of the RGB location
on the age and on Z.
\begin{figure} \centerline{
\psfig{figure=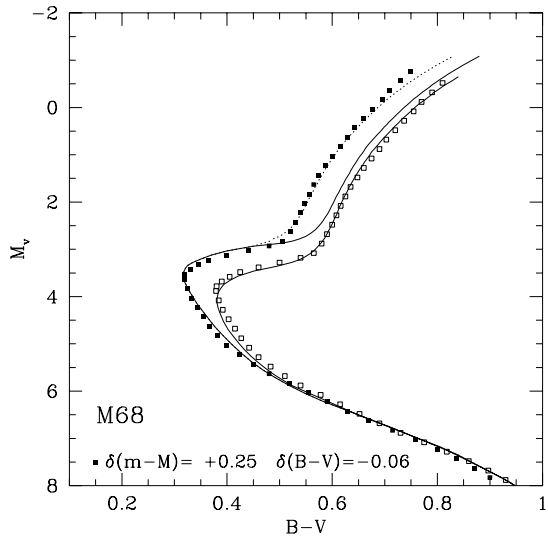,height=7.5cm} } \caption{A schematic view of the
RGB location of a metal poor GC (M68), if a `short' distance scale (age=16Gyr
open squares) or a `long' one (age=12Gyr, full squares) is adopted. The RGB
fit requires for the long distance scale a more efficient convection than the
short one. In this sense, all set of tracks have their convection treatment
`calibrated' on the distance scale. }
\label{gc1} \end{figure}
\subsection{ The `hot' boundary of the RGB location}
First of all, models for population synthesis must reproduce the
location of the most metal poor GCs. As the convection theories presently
available can not predict the RGB $T_{\rm eff}$,
our models must choose an appropriate calibration (in fact, we do not
fit $T_{\rm eff}$'s, but colors, so that we also need to know the
correspondence between colors and $T_{\rm eff}$). It is easy to convince
ourselves that the location of old metal poor giants in the magnitude --
color
plane depends in a subtle way on the adopted distance scale (or age). Figure
\ref{gc1} shows in a schematic way that distance scales differing by 0.25mag
for the GC M68 result either in a fit with 16Gyr (on the right, open squares)
or with 12 Gyr ---after adjusting the colors of the diagram by 0.06mag. But,
if the stellar models were able to fit the giants location at an age of
16Gyr, the same models provide a RGB {\sl too cool} for the younger age. The
obvious way to have again the RGB fit is to {\sl increase the efficiency of
convection}, e.g. by increasing the $\alpha$ in the stellar models. In this
way we see that the left boundary of the old, metal poor RGBs is ``fixed by
the distance scale of the metal poor GCs". In the case exemplified, the
uncertainty amounts to $\sim 0.18$mag in the average $B-V$\ color of the RGB.
Actually, it is less severe, as there are many metal poor GCs which can be
used as calibrators, opacities are relatively well known just because Z is
very low, and it is not unrealistic to agree to the common belief that the
metal poorest GCs are all coeval.

\subsection{The `cool' boundary of the RGB location}
Increasing the metallicity, the RGB location shifts to lower $T_{\rm eff}$'s
(redder colors). As we can not trust blindly the dependence Z -- $T_{\rm
eff}$\ of our models, even if we have well calibrated them at low Z, we must
be able to define and make use of a `right boundary' to the RGB location.
This task is not yet well exploited, due to the
following problems:
\begin{figure}
\centerline{
      \psfig{figure=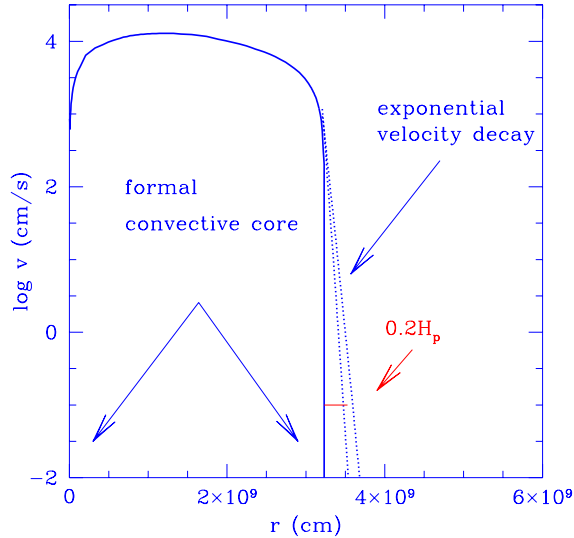,height=7.5cm}
}
\caption{The average velocity inside the core of a MS 4.5M$_\odot$\ star is
shown, computed according to the FST model by Canuto, Goldman and Mazzitelli
1996. The dashed lines show the extrapolation of the convective velocities
assumed in the models with non instantaneous mixing and overshooting. The
sharper decline corresponds to the overshooting scale $ov$=0.02, the other
to $ov$=0.03.} \label{fig2}
\end{figure}

1. Calibrators for higher metallicity become less and less secure, as
the `observational' metallicity is not certain, particularly if we have also
to consider possible $\alpha$--elements enhancements. Further, an age spread
is possible. Let us consider that ages from 8 to 14 Gyr appear in the recent
literature for the GC 47Tuc (Salaris \& Weiss 1998; Vazdekis et al. 2001),
which certainly is not extreme in metallicity ([Fe/H]$\sim-0.9$ with
$[\alpha/Fe] \sim 0.4$). 

2. The $T_{\rm eff}$'s of the RGB become quite low: below $\sim
4000$K, triatomic molecules ({\it in primis} H$_2$O) must be included in the
opacity computations --remember that they are not included in the widely
used Kurucz's libraries--; model atmospheres must no longer rely on simple
grey integration to be used as boundary conditions to the stellar models.

3. Even if we knew how to deal with convection, the computation of the
overadiabatic gradient requires knowledge also of the radiative gradient:
consequently, we need a detailed opacity treatment, also at large optical
depth.

My conclusion is that we can not state that we know the RGB location for
`large' metallicities well enough to trust synthetic evolution
models, in order to infer in detail the age of compact far stellar
systems from their integrated colors. In particular, any correlation
age -- Z derived from integrated colors must be well checked in
order to understand whether it is not mainly a result of the assumed
theoretical dependence of the RGB location on the metallicity.
\subsection{The `hot bottom' of convective envelopes}
The efficiency of convection determines also the structure of a distinct
group of stars: the very luminous Asymptotic Giant Branch (AGB) stars, which
are known to reach, at the bottom of their convective
envelope, temperatures which allow nuclear burning. There are today
many models which
successfully predict, e.g., the formation of Lithium in the envelopes of the
luminous AGBs (Sackmann and Boothroyd 1992, Mazzitelli, D'Antona \& Ventura
1999, Bl\"ocker, Herwig \& Driebe 2000) in agreement with their
observational counterparts of the Magellanic Clouds (Smith et al. 1995).
The {\sl minimum} mass which suffers `hot bottom burning' (HBB) depends on
the metallicity and on the convection efficiency. A possible way to study the
convection efficiency is then to search for Lithium rich AGBs in cluster of
known age (and thus, in principle, of known evolving mass).
Such stars have been recently found in two young clusters of the LMC: one
in NGC 1866 and one in NGC 2031 (Maceroni et al. 2001).
\section{Convection model: mixing beyond the convective boundary}
A second aspect of convection in stars is the chemical mixing
associated with convection, and in particular the mixing which extends
beyond the formal convective boundaries, defined in most cases by the
Schwarzschild criterion. This problem is generally treated in the context of
an `overshooting' problem, although there are other physical mechanisms
which can result in an extension of the stellar convective cores larger than
predicted by standard stellar models (e.g. rotation, Maeder \& Meynet 2000).
In past years, an inadequate knowledge of opacities was at the basis of
prediction of convective cores smaller than required by the observations, as
shown at the introduction OPAL opacities (Roger \& Iglesias 1992, see e.g.
Stothers \& Chin 1992).
I wish to discuss only `overshooting' here, that is the extramixing due to
the fact that the convective eddies arrive at the Schwarzschild boundary with
zero acceleration, but with a finite velocity which can allow them to go
beyond it. A self-consistent treatment of this problem has been until now
beyond
solving possibilities, as a non local model for convection is necessary.
The analytic approaches have been unfruitful for most cases of
astrophysical interest (that is: when extended convective regions are
present, but see, e.g., Kupka 1999 for the description of thin convective
envelopes), and the LES can not yet deal with the description of the
turbulence at all the scales present in stars. Consequently, stellar model
still adopt a local description of turbulence (like the MLT or FST) and a
fully parametric approach for the overshooting.

In addition, most stellar models adopt a very rough approach to
describe overshooting: it is assumed that the convective
mixing proceeds for a given amount beyond the formal convective border. This
is generally parametrized in fractions of $H_p$\ at the border. A value of
$\sim 0.2H_p$\ seems plausible and is adequate to fit, e.g., the
observed thickness of the main sequence. A dependence of the degree of
overshoot on the stellar mass (larger overshooting the larger is the mass)
seems however to be required. This enlarges the number of parameters we need
to describe convection and does not allow us to make predictions.
\begin{figure}
\centerline{
      \psfig{figure=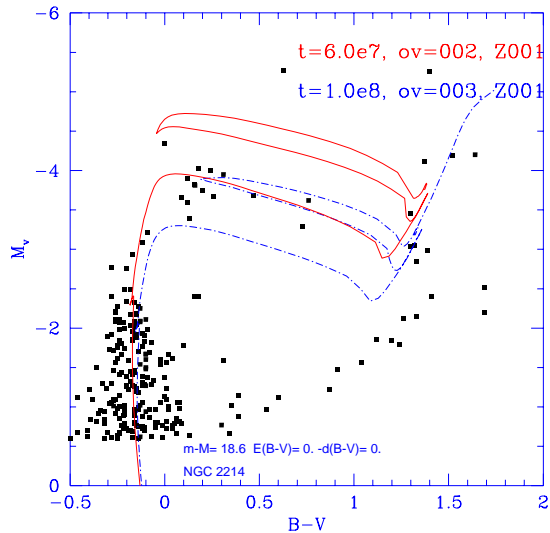,height=7.5cm}}
\caption{HR diagram of the LMC cluster NGC 2214, and two isochrone `fits':
only the older age is realistic, based on the expected number of clump
stars.}
\label{fig4}
\end{figure}
Deng et al. (1996) introduced non-instantaneous mixing assuming an
exponential decay of the convective velocities outside the formal convective
region. A similar approach has been introduced by Ventura et al. (1998).
Figure \ref{fig2} shows the convective velocities in the core
of a 4.5$M_\odot$\ star according to the FST treatment (Canuto et al. 1996).
As the convection model is local, the velocity decreases to zero both at the
inner and outer boundaries. The spurious decrease at r=0 is
simply ignored in the model. At the outer boundary, we
extrapolate the velocity before its sharp decrease, and impose that the eddy
velocities decay exponentially, with a given e--folding distance ($ov$)
(details in Ventura et al. 1998). A 0.2$H_p$\ overshoot is also shown for
comparison. Mixing inside the convective
and overshooting regions is solved, together with the nuclear evolution
matrix, as a time dependent diffusion process (diffusive mixing).
\begin{figure}
\centerline{\hbox{
      \psfig{figure=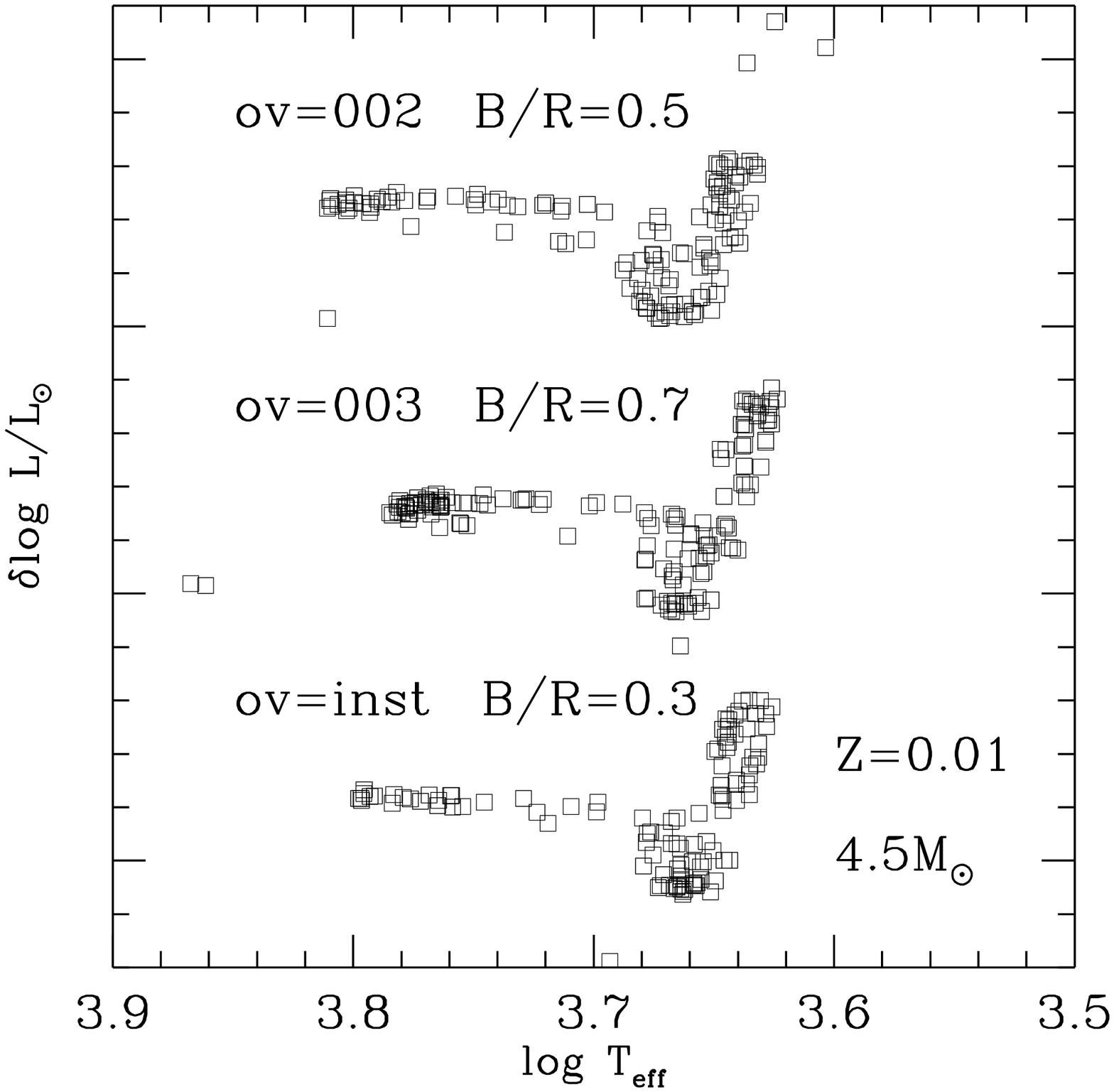,height=6.3cm}
      \psfig{figure=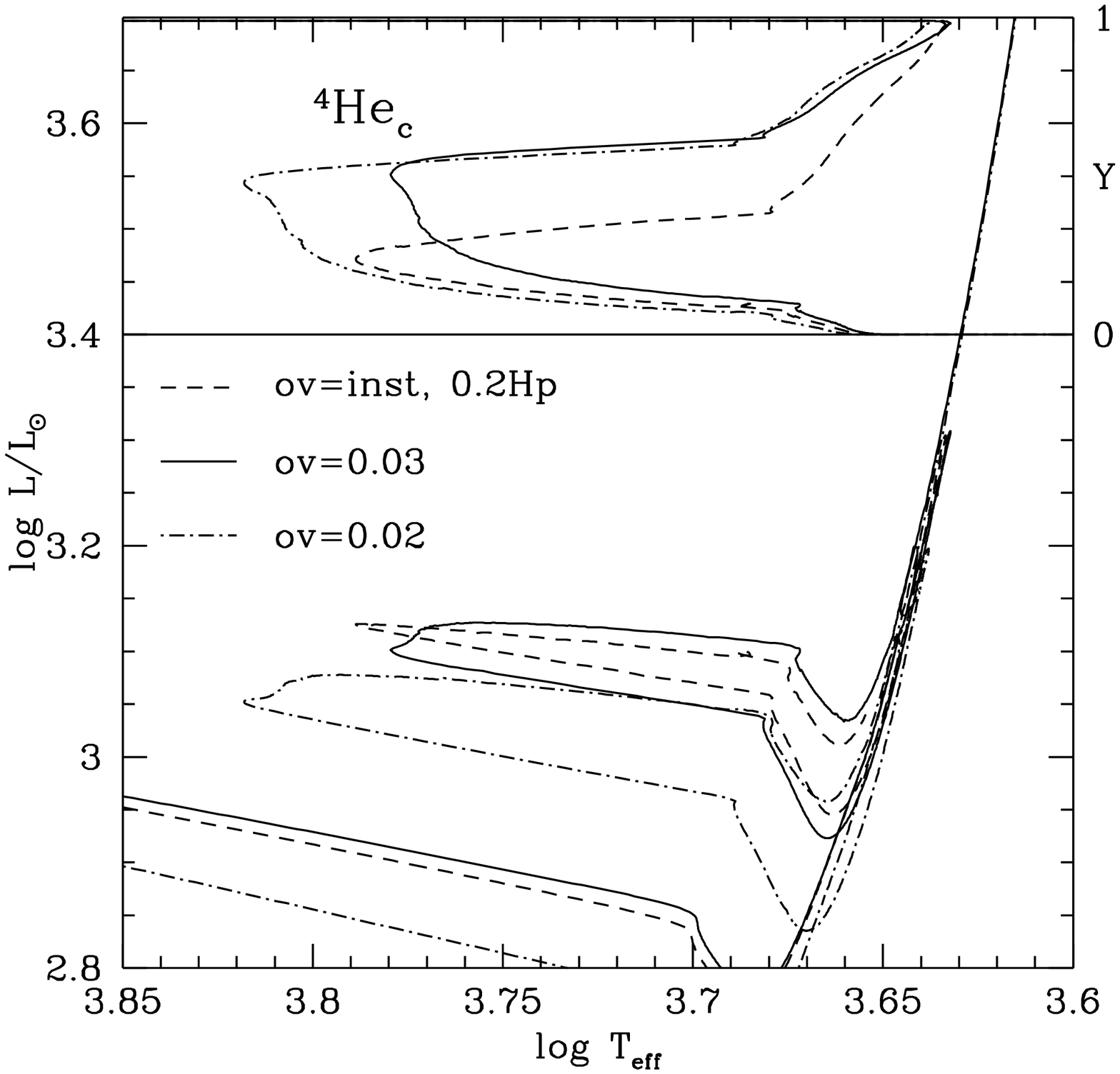,height=6.3cm}
}}
\caption{On the right the evolutionary tracks for 4.5M$_\odot$\ and
different overshooting treatment are shown, displaying on top the core
helium abundance Y along the evolution. The left figure shows simulations
based on the tracks: the number ratio of clump stars (B) to RGB stars (R)
predicted by each model is given. The observed ratio is $\sim 0.9$. The
$ov$=inst model predicts very few stars in the blue loop.}
\label{fig3}
\end{figure}
This modality of mixing is substantially different from the instantaneous
mixing approximation, and the models must be compared to observations to
understand whether they improve the agreement. This is being done by Tosi et
al. (2001) with a synthetic HR diagram approach, superior to the
plain comparison of isochrones: this latter in fact does not take into
account the lifetimes in the different evolutionary phases. As an example,
Figure \ref{fig4} shows two isochrones superimposed to the HR diagram of the
young cluster NGC 2214 in the LMC. For this cluster, Subramaniam \& Sagar
(1995) derive an age of 40 -- 60Myr, but the stars which seem to be post-MS
when compared to the 60Myr isochrone, may actually be in the clump stage of a
100Myr isochrone, as it can be shown by comparing with synthetic HR
diagrams.
\subsection{The clump of NGC 1866}
A detailed study of the core--He burning phase with diffusive mixing and
overshooting will be given by Ventura et al. (in preparation).
The extension and duration of the blue loop depends on the fine balancing
between three main parameters: 1) the intensity and phase of core
He--burning; 2) the ratio between the efficiency of the CNO H--burning
shell and the core He--burning; 3) the opacity in the surface layers (Iben
1967). These parameters are influenced by the modelling of convection, both
through the temperature gradient (3) and through overshooting in this phase
and in the preceding ones (1 and 2). We can find very fast and/or small
loops, or extended and long lasting loops depending on the inputs. Figure
\ref{fig3} exemplifies the results obtained for different convection models
in the case of a 4.5M$_\odot$\ evolution, which well describes the evolution
in the cluster NGC 1866. The degree of overshooting and its modalities
(diffusive versus instantaneous) determine the location of the clump and the
ratio between ``blue" (B) and ``red" (R) stars. The instantaneous treatment
of overshooting in this case provides the worst performance.
\begin{figure}
\centerline{\hbox{
      \psfig{figure=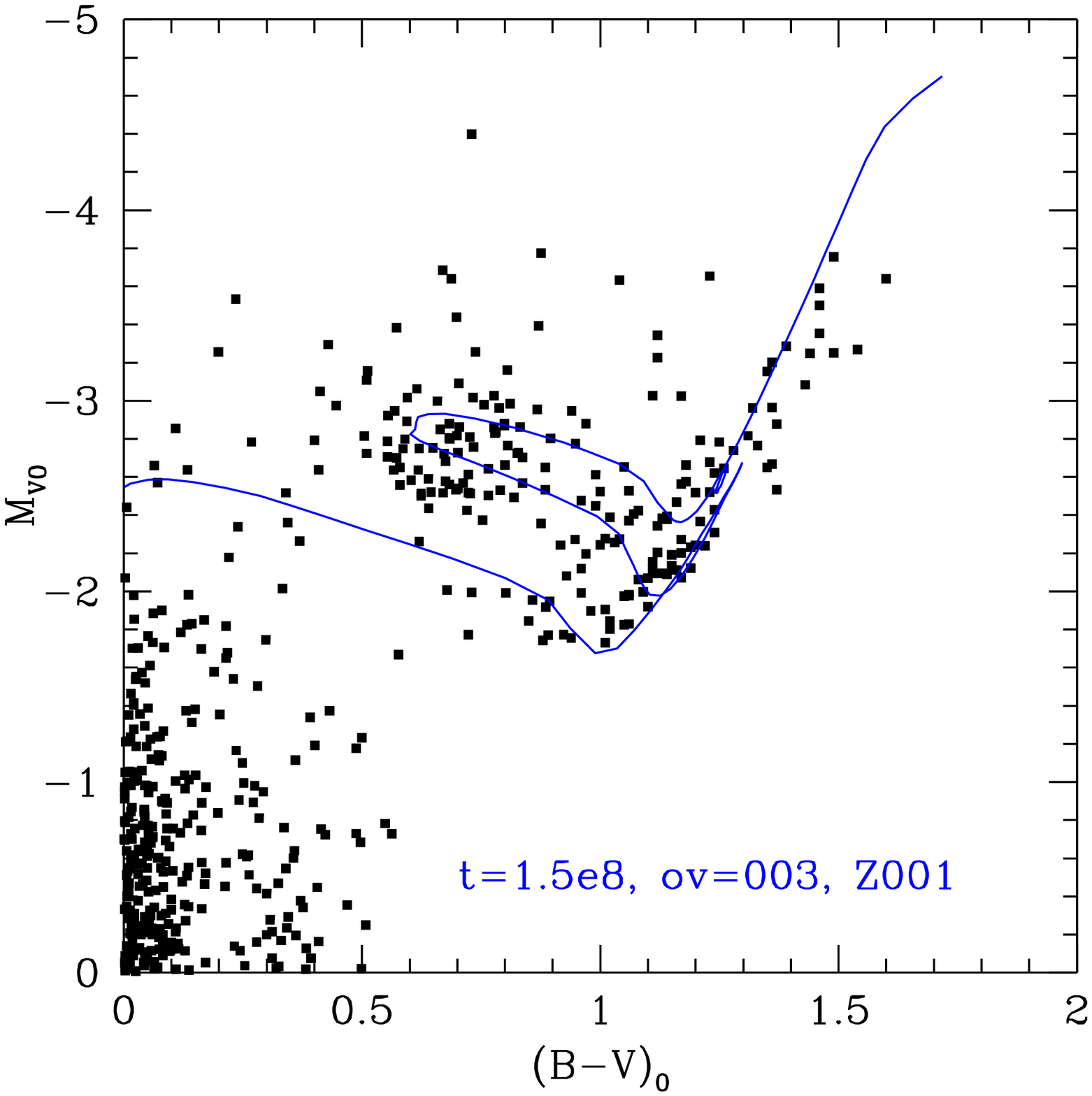,height=6.cm}
      \psfig{figure=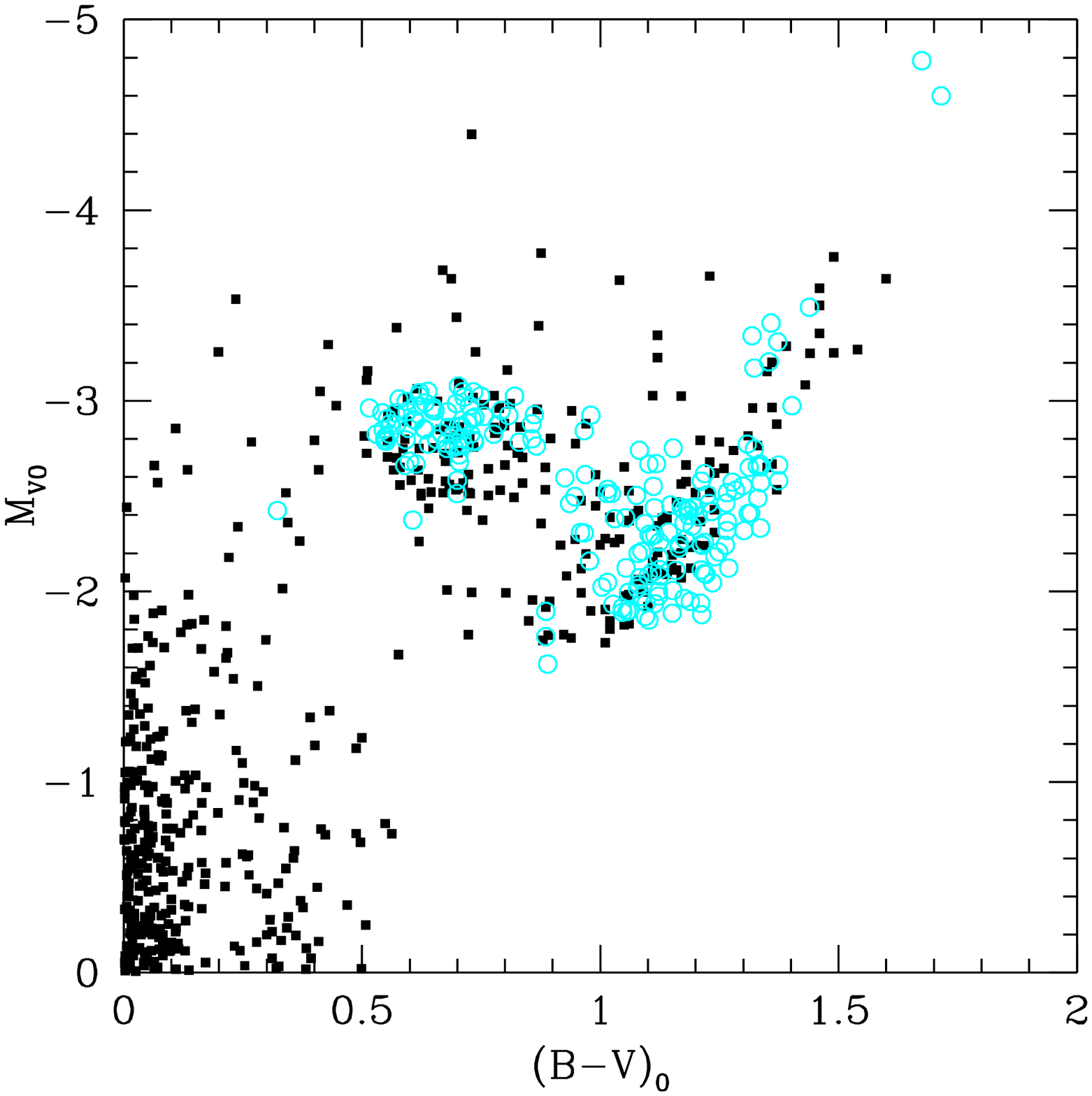,height=6.cm}
}}
\caption {In the left figure, an isochrone of 150Myr is superimposed to the
HR diagram of the clump stars of NGC 1866 (Testa et al. 1999) (distance
modulus 18.6mag). On the right, a simulation of the cluster population is
superimposed (open circles).}
\label{fig5}
\end{figure}
\\ \\ In conclusion, the HR diagram of NGC 1866, and in
particular its clump, is well reproduced
with the models with $ov$=0.03, for an age of about 1.5$\times 10^8$yr.
Figure \ref{fig5} shows the comparison of the HR diagram by Testa et al.
(1999) with the isochrone (left) and with the synthetic model (right). At
this age, a mass of $\simeq 4.5M_\odot$\ is evolving in the AGB, and for this
mass our models predict HBB and the production of Lithium, confirmed to be
present in one AGB star of the cluster by Maceroni et al. (2001).
Although still very parametric, these models seem to offer very detailed
performances in the comparisons with observations, and allow us a step ahead
in the understanding of stellar clusters.

\acknowledgements I thank very much Eva Grebel and Doug Geisler for inviting
me to deliver this talk and for their organizational effort (resulting in a
very good Symposium and a spectacular environment).

\section*{Discussion}

{\it H. Lamers:\, }
The occurrence  of  red or blue horizontal branches is most likely related to
the amount of envelope mass that remained on the stars after the Helium
flash. Is it possible to relate the observed occurence of red or blue
horizontal branches of globular clusters to any expected process that may
have influenced the mass loss rate in the phases BEFORE  the Helium flash.
\\ \\

\noindent
{\it F. D'Antona:\, }
Yes indeed. Your question rises again the problem of the `second parameter'
of Globular clusters. Mass loss is in fact the other important issue of
``macrophysics" which we do not know how to manage from first principles.
In GC stars, mass is probably lost not only at the helium
flash, but also during the most luminous phases of the RGB evolution.
The H layer left on the star at the HB phase depends of course also on the
total mass which we see in evolution, and thus on the age, which, in
principle, can affect the stars distribution in the HB. However, there are
many cases in which we know that age {\it is not} the second parameter, and
`environmental' parameters affect the mass loss: e.g. the dynamical
interactions among stars due to the stellar densities play a role, as nicely
shown also here, in the poster by Catelan, Rood and Ferraro. The HB
distribution of stars in M3 depends on the distance from the cluster center,
the innermost cluster regions having {\sl many more} blue HB stars than the
outermost ones. This can only be explained if the stars in the denser regions
have lost more mass on the RGB. \\ \\

\noindent
{\it G. Harris:\,}
What are the possibile ways of changing the stellar models in order to reach
the observed number ratio of clump stars versus red giant stars in NGC 1866?
\\ \\

\noindent
{\it D'Antona:\,}
In the light of my talk, I stress that we can probably get a decent fit for
the observed cluster, but this also means that theoretical models still do
not have a satisfactory predictive power. In particular, our models for NGC
1866 reach a ratio of clump stars to RGB stars of B/R$\sim 0.7$, while the
observed value is even larger ($\sim 0.9$). I wished to stress that the clump
is very dependent on the details of convection at several stages: on the one
hand it depends on the maximum extension of the H--envelope convection, which
leaves a chemical CNO discontinuity crossed by the H--burning shell during
the core He--burning stage, ultimately producing the loop. `Non canonical'
mixing mechanisms may also affect the chemical discontinuity. On the other
hand, the ratio B/R depends on the He-core burning stage (figure 4), which in
part depends on the way we feed with fresh He the core. Instantaneous mixing
and overshooting seems to give too small B/R ($\sim 0.3$), as the blue loop
occurs when the core He is already very small, but a precise tuning sure
requires that we explore many other parameters in addition to the role of
diffusive mixing (I wish to remember here, e.g., your exam of the role of
metallicity, Harris and Deupree 1976). Paolo Ventura is preparing a
discussion of this problem.

\end{document}